\documentclass{article}

\usepackage{academicons,url}
\usepackage{graphicx,multirow}
\usepackage{amsmath,amsfonts}
\usepackage{xcolor,colortbl}
\usepackage{url,siunitx,bbm}
\usepackage[export]{adjustbox}
\usepackage[utf8]{inputenc}
\usepackage[normalem]{ulem}
\usepackage[compact]{titlesec}
\usepackage{caption,subcaption}

\newtheorem{definition}{Definition}
\newtheorem{theorem}{Theorem}
\newtheorem{corollary}{Corollary}[theorem]
\newtheorem{lemma}{Lemma}[theorem]
\newtheorem{example}{Example}

\newcommand*\di{\mathop{}\!\mathrm{d}}

\title{Games in the Time of COVID-19: \\
    Promoting Mechanism Design for Pandemic Response}

\author{\textbf{Balázs Pejó and Gergely Biczók} \\
Laboratory of Cryptography and System Security \\
Department of Networked Systems and Services \\
Faculty of Electrical Engineering and Informatics \\
Budapest University of Technology and Economics \\
\textit{surname}@crysys.hu}

\date{}
 
\begin{document}

\maketitle

\begin{abstract}
Most governments employ a set of quasi-standard measures to fight COVID-19 including wearing masks, social distancing, virus testing, contact tracing, and vaccination. However, combining these measures into an efficient holistic pandemic response instrument is even more involved than anticipated. We argue that some non-trivial factors behind the varying effectiveness of these measures are selfish decision making and the differing national implementations of the response mechanism. In this paper, through simple games, we show the effect of individual incentives on the decisions made with respect to mask wearing, social distancing and vaccination, and how these may result in sub-optimal outcomes. We also demonstrate the responsibility of national authorities in designing these games properly regarding data transparency, the chosen policies and their influence on the preferred outcome. We promote a mechanism design approach: it is in the best interest of every government to carefully balance social good and response costs when implementing their respective pandemic response mechanism; moreover, there is no one-size-fits-all solution when designing an effective solution.
\end{abstract}

\section{Introduction}

The current coronavirus pandemic is pushing individuals, businesses and governments to the limit. Even with the recently emerged hope of rapidly developed vaccines, people still suffer owing to restricted mobility, social life and income, complete business sectors face an almost 100\% drop in revenue, and governments are scrambling to find out when and how to impose and remove restrictions. In fact, COVID-19 has turned the whole planet into a ``living lab'' for human and social behavior where feedback on response measures employed is only delayed by around two weeks (the incubation period). From the 24/7 media coverage, all of us have been introduced to a set of quasi-standard measures introduced by national and local authorities, including wearing masks, social distancing, virus testing, contact tracing, vaccination, and so on. It is also clear that different countries have had different levels of success employing these measures as evidenced by the varying normalized death tolls and confirmed cases\footnote{Johns Hopkins Coronavirus Resource Center. \url{https://coronavirus.jhu.edu/map.html}}. 

We believe that apart from the intuitive (e.g., genetic differences, medical infrastructure availability, hesitancy, etc.), there are two significant factors that have not received enough attention. First, the \emph{individual incentives} of citizens, e.g., ``is it worth more for me to stay home than to meet my friend?'', have a significant say in every decision situation. While some of those incentives can be inherent to personality type, clearly, there is a non-negligible rational aspect to it, where individuals are looking to maximize their own utility. Second, countries have differed in their specific \emph{implementation} of response measures, e.g., whether they have been distributing free masks (affecting the efficacy of mask wearing in case of equipment shortage), regulating the amount of accepted vaccines (affecting the speed of reaching herd-immunity), or providing extra unemployment benefits (affecting the likelihood of proper self-imposed social distancing). Framing pandemic response as a mechanism design problem, i.e., architecting a complex response mechanism with a preferred outcome in mind, can shed light on these factors; what's more, it has the potential to help authorities (mechanism designers) fight the pandemic efficiently. The objective of this paper is to show that both individual incentives and the actual design and implementation of the holistic pandemic response mechanism can have a major effect on how this pandemic plays out.

\paragraph{Contribution. }In this paper we model decision situations during a pandemic with game theory where participants are rational, and the proper design of the games could be the difference between life and death. We extend and improve on our initial paper~\cite{pejo2020corona}. Our main contribution is two-fold. 
First, regarding decisions on wearing a mask, we show that i) the equilibrium outcome is not socially optimal under full information, and ii) when the status of the players are unknown the equilibrium is not to wear a mask for a wide range of parameters. Furthermore, for social distancing, using current COVID-19 statistics we show that i) going out is only rational when it corresponds to either a huge benefit or staying home results in a significant loss, and ii) we determine the optimal duration (or meeting size) of such an out-of-home activity. 
Moreover, we study various aspects of two hypothetical vaccines from the individual point of view, including i) the efficiency of the protection the vaccine provides, ii) the duration of the immunity granted, iii) the temporal availability when the individual can get vaccinated, and iv) the potential side-effect. Based on these we build and analyze several decision models (i.e., optimization problems) for individuals, and also show when early or late vaccination is the likely outcome in the population. 
Second, we take a look at pandemic response from a mechanism design perspective, and demonstrate that i) different government policies influence the outcome of these games profoundly, and ii) standalone response measures (sub-mechanisms) are interdependent. Specifically, we discuss i) how the quality and the truthfulness of public reports on pandemic data impacts individual decision making, ii) how contact tracing enables targeted testing which in turn reduces the uncertainty in individual decision making regarding both social distancing and wearing masks, and iii) which factors an approximately optimal vaccination policy has to consider. We recommend that governments treat pandemic response as a mechanism design problem when weighing response costs versus the social good.

\paragraph{Organization. }
The remaining of the paper is structured as follows. 
In Section~\ref{sec:related} we briefly describe related work while in Section~\ref{sec:pre} we recap some basics of game theory. 
In Section~\ref{sec:mask} we develop and analyze the two player Mask Game with uncertainty and efficiency, and also introduce the multi-player extension of the basic model. 
In Section~\ref{sec:sd} we develops and analyze the Distancing Game which includes the effects of meeting duration (or size). 
In Section~\ref{sec:vacc} we develop and analyze several vaccination-based decision models focusing on various aspects such as efficiency and duration and present the Availability Game to facilitate the choices of individuals between various vaccines. 
In Section~\ref{sec:mech} we frame pandemic response as a mechanism design problem using the games introduced above. 
Finally, in Section~\ref{sec:con} we outline future work and concludes the paper.

\section{Related Work}
\label{sec:related}

In this section we review some well-known and/or recent epidemic response mechanisms and game-theoretic works in relation to masks, social distancing, vaccination, and pandemics in general. A comprehensive systematic literature review on COVID-19 can be found in~\cite{rajendran2020systematic}.

\subsection{Pandemic Response Models}

COVID-19 has been modelled using different epidemic models, e.g., using SIR~\cite{kiamari2020covid}, SEIQR~\cite{zhang2020impact}, and SIDARTHE~\cite{giordano2020modelling}. Concerning the parameters of these epidemiological models the authors in~\cite{wang2020sensitivity} performed a global sensitivity analysis of the model parameters to see their effect on the time when the number of cases are peaking. 

Besides the model, the seed data instantiating the model may be imperfect as well; this problem was studied in~\cite{hong2020estimation}. A similar issue was also tackled in~\cite{thakur2020covid,aboubakr2020improving} where the authors developed a platform allowing informed decision making, and proposed an improved epidemic case reporting mechanism, respectively.

An orthogonal extension of these models is proposed in~\cite{santosh2020covid} that discusses how factors such as hospital capacity, test capacity, demographics, population density, vulnerable population segments and income could be integrated into these models. In contrast to the previous models, the one in~\cite{lagos2020games} takes into consideration the networked structure of human interconnections and the locality of interactions, without attempting a mean-field approach. Spatial networks were also utilized in~\cite{medo2020contact} for contact tracing. Another paper~\cite{agarwal2020infection} introduces an infection risk score that provides an estimate of the infection risk stemming from human contacts using a mobile application. 

\subsection{Game-Theoretic Models}

We briefly review some related research efforts in the intersection of epidemics and game theory. For a comprehensive survey we refer the reader to~\cite{chang2019game}.

Behavioral changes of people caused by a pandemic and, specifically, COVID-19 was studied in~\cite{poletti2012risk,brune2020evolutionary}, respectively. Others focused on the mobility habits of people traveling between areas affected unevenly by the disease~\cite{zhao2018strategic}. The authors of~\cite{ozkaya2021effects} took a closer look through the lens of game theory on the effect of self-quarantine on virus spreading. In~\cite{bairagi2020controlling} an optimization problem was formalized by accommodating both isolation and social distancing. The impact of social distancing was also studied in~\cite{choi2020optimal} in combination with vaccines. 

Several other studies also focused on how the availability of vaccines affects human behaviour. Vaccination order based on spatial networks was studied in~\cite{piraveenan2020optimal}, while the model introduced in~\cite{bhattacharyya2011wait} studied personal vaccination preferences. In the latter, the authors concluded that vaccine delayers relied on herd immunity and vaccine safety information generated by early vaccinators. Consequently, the Nash Equilibrium was ``wait and see''. Another study concerning this vaccination dilemma proposed a model with incentives for individuals to choose the prevention strategy according to risks and expenses in the epidemic campaign~\cite{bauch2004vaccination}. Similarly, researchers in~\cite{van2008self} showed the optimal use of anti-viral treatment by individuals when they took into account the direct and indirect costs of treatment. The game-theoretic model in~\cite{sun2009selfish} focused on the various level of drug stockpiles in different countries, and found controversial results: sometimes there was an optimal solution with a central planner (such as the WHO), which improved on the decentralized equilibrium, but other times the central planner's solution (minimizing the number of infected persons globally) required some countries to sacrifice part of their population. 

Game theory was also used for scarce and personal equipment in~\cite{salarpour2021multicountry,abedrabboh2021game}, respectively. The exact dynamics of demand and supply for medical resources at different phases of a pandemic was also studied~\cite{chen2020pandemic}. Being able to correctly forecast such dynamics would provide a quantitative basis for mechanism designers (e.g., decision makers of healthcare systems) to understand the potential imbalance of supply and demand. The authors extended the concepts of reserving and capital management in the classical insurance literature, and aimed to provide a framework for quantifying and assessing pandemic risk, and developed optimal strategies for stockpiling spatiotemporal resources.

The Centers for Disease Control and Prevention created a policy review of social distancing measures for pandemic influenza in non-healthcare settings~\cite{fong2020nonpharmaceutical}. They identified measures to reduce community influenza transmission such as isolating the sick, tracing contacts, quarantining exposed people, closing down school, changing workplace habits, avoiding crowds, and restricting movement. The impact of several of these (and wearing masks) was studied in~\cite{silva2020covid},  where authors used agent-based modelling for the pandemic, simulating actions of people, businesses and the government. Other researchers demonstrated that early school and workplace closures, and the restriction of international travel are independently associated with reduced national COVID-19 mortality~\cite{papadopoulos2020impact}. On the other hand, lock-down procedures could have a devastating impact on the economy. This was studied in~\cite{chao2020simplified} with a modified SIR model and time-dependent infection rate. The authors found that, surprisingly, in spite of the economic cost of the loss of workforce and incurred medical expenses, the optimum point for the entire course of the pandemic is to keep the strict lock-down as long as possible.

As detailed above, related work has mostly studied narrowly focused specifics of epidemic modelling such as the intricate behaviour of individuals in relation with vaccines, or the preferred actions of mechanism designers such as healthcare system operators. In contrast, our work takes a step back, and focuses on the big picture: we model decision situations during a pandemic as games with rational participants, and promote the proper design of these games. We highlight the responsibility of mechanism designers such as national authorities in constructing these games properly with adequately chosen policies, taking into account their interdependent nature. 

\section{Preliminaries}
\label{sec:pre}

In this section we shortly elaborate on the main game theoretical notions used in this paper, to facilitate the conceptual understanding of the implications of our results. 

Game theory~\cite{harsanyi1988general} is ``the study of mathematical models of conflict between intelligent, rational decision makers''. Almost any multi-party interaction can be modeled as a game. In relation to COVID-19, decision makers could be individuals (e.g., whether to wear a mask), municipalities (e.g., whether to enforce wide-range testing within the city), governments (e.g., whether to apply contact tracing within the country), or companies (e.g., whether to apply social distancing within the workplace). Potential decisions are referred to as strategies; decision makers (players) choose their strategies rationally: i) they maximize their own utility, and ii) they do not make mistakes. Rationality is a standard assumption in game theory; in fact, it enables the succinct evaluation of non-cooperative games, yielding valuable insights even from simple models. 

Note that rational (in a game-theoretical context) does not necessarily mean fully and objectively informed, i.e., individuals will make their decisions based on the \emph{perceived} utility of their actions. Such a decision can even go against scientifically proven best practices, resulting in refusing vaccination or partying carelessly.
Naturally, more realistic behavioral modelling (e.g., bounded rationality, unpredictability and a large number of proven behavioral biases~\cite{ariely2008predictably}) delves deeper into the human decision-making process. However, the simple decision models in this paper serve more of a demonstrative purpose, illustrating i) how (selfish) individual decisions perturb society-level behavior, and ii) how central mechanism design decisions influence the outcome of such models. Consequently, the direct practical applicability of the outcome of these simplified games is rather limited; however, they promote a mechanism design mindset which, indeed, should be the norm for complex pandemic response. As to the behavioral economics models of medical scenarios, clearly out of scope for us this time, we refer the interested reader to, e.g., ~\cite{plonsky2021underweighting} for the biased expectation concerning rare events in social interactions such as distancing, and~\cite{roth2020value} for the importance of alert systems and gentle rule enforcement~\cite{erev2020complacency} in pandemic mechanism design.

The Nash Equilibrium (NE)---arguably the most famous solution concept---is a set of strategies where each player's strategy is a best response strategy. This means every player makes the best/optimal decision for itself as long as the others' choices remain unchanged. NE provides a way of predicting what will happen if several entities are making decisions at the same time where the outcome also depends on the decisions of the others. The existence of a NE means that no player will gain more by unilaterally changing its strategy at this unique state. Another game-theoretic concept is the Social Optimum (SO), which is a set of strategies that maximizes social welfare. Note, that despite the fact that no one can do better by changing strategy, NEs are not necessarily Social Optima (we refer the reader to the famous example of the Prisoner's Dilemma~\cite{harsanyi1988general}). In fact, it is well-studied in game theory how much a distributed outcome (NE) is worse than a centrally planed social optimum; this ratio is captured by the Price of Anarchy (PoA, \cite{koutsoupias1999worst})

If one knows the NE they prefer as the outcome of a game, e.g., everybody wearing a mask, and they have the power to instantiate the game accordingly, i.e., fixing the structure, game flow and any free parameters, then we talk about mechanism design~\cite{mas1995microeconomic}. In a way, mechanism design is the inverse of game theory; although a significant share of efforts within this field deals with auctions, mechanism design is a much broader term applicable to any multi-stakeholder mechanism, (e.g., optimal organ matching for transplantation, school-student allocation or, in fact, pandemic response), aimed at achieving a preferred steady state result. 

\section{The Mask Game}
\label{sec:mask}

Probably the most visible consequence of COVID-19 are masks: before, their usage was mostly limited to some Asian countries, hospitals, constructions and banks (in case of a robbery). Due to the coronavirus pandemic, an unprecedented spreading of mask-wearing can be seen around the globe. Policies have been implemented to enforce their usage in some places, but in general, it has been up to the individuals to decide whether to wear a mask or not, based on their own risk assessment. In this section, we model this decision situation via game theory. We assume that there are several types of masks, providing different levels of protection. 

\begin{itemize}
    \item \textbf{No} Mask corresponds to the behavior of using no masks during the COVID-19 (or any) pandemic. Its cost is consequently 0; however, it does not offer any protection against the virus. 
    \item \textbf{Out} Mask is the most widely used mask (e.g., cloth mask or surgical mask\footnote{FFP2 masks are mandated in some countries (e.g., Germany) for certain activities (e.g., using public transportation)}). They are meant to protect the environment of the individual using it. They work by filtering out droplets when coughing, sneezing or simply talking, therefore they limit the spreading of the virus. They do not protect the wearer itself against an airborne virus. The cost of deciding for this protection type is noted as $C_{out}>0$.
    \item \textbf{In} Mask is the most protective prevention gear designed for medical professionals (e.g., FFP2 or FFP3 mask with valves). Valves make it easier to wear the mask for a sustained period of time, and prevent condensation inside the mask. They filter out airborne viruses while breathing in; however, the valved design means they do not filter the while air breathing out. Note that CDC guidelines\footnote{Centers for Disease Control and Prevention. \url{https://www.cdc.gov/coronavirus/2019-ncov/prevent-getting-sick/prevention.html}} recommend using a cloth/surgical mask for the general public, while valved masks are only recommended for medical personnel in direct contact with infected individuals. The cost of this protection type is $C_{in}>>C_{out}$.
\end{itemize}

Besides which mask they use (i.e., the available strategies), the players are either susceptible or infected\footnote{We simplify the well-known SIR model~\cite{diekmann2000mathematical} since in case of COVID-19 it is currently unclear if and for how long an individual is resistant after recovery.}. The latter has some undesired effects; hence, we model it by adding a cost $C_i$ to these players' utility (which is magnitudes higher than even $C_{in}$, i.e., $C_i>>C_{in}>>C_{out}$). We summarize all the parameters and variables used for the Mask Game in Table \ref{tab:parMASK}. Using these states and masks, we can present the basic game's payoffs where two players with known health status meet, and decide which mask to use. 

\begin{definition}
    The basic Mask Game is a tuple $\langle\mathcal{N},\Sigma,\mathcal{U}\rangle$, where the set of players is $\mathcal{N}=\{1,2\}$ and their actions are $\Sigma=\{\textbf{no}, \textbf{in}, \textbf{out}\}$. The utility functions $\mathcal{U}=\{u_1,u_2\}$ are presented as a cost matrix in Table \ref{tab:mask_game_payoffBOTH} and \ref{tab:mask_game_payoffONE}. In details, Table \ref{tab:mask_game_payoffBOTH} corresponds to the case when both players are susceptible, while Table \ref{tab:mask_game_payoffONE} corresponds to the case when one player is infected while the other is susceptible. Note that when both players are infected, the payoff matrix would be as when both are susceptible, with an additive constant cost $C_i$.
\end{definition}

\begin{table}[t]
\centering
    
    \begin{tabular}{c|l}
        Variable & Meaning \\
        \hline
        $C_{out}$ & Cost of playing \textbf{out} \\
        $C_{in}$ & Cost of playing \textbf{in} \\
        $C_i$ & Cost of being infected \\
        \hline
        $C_{use}$ & Cost of playing \textbf{use} \\
        $\rho$ & Probability of being infected \\
        \hline
        $a$ & Protection Efficiency \\
        $b$ & Spread Prevention Efficiency \\
        \hline
        $g$ & Meeting Size \\
    \end{tabular}
    \vspace{0.1cm}
    \caption{Parameters of the various Mask Games}
    \label{tab:parMASK}
    
    \begin{tabular}{c|ccc}
        & \textbf{no} & \textbf{out} & \textbf{in}\\
        \hline
        \textbf{no} & $[0,0]$ & $[0,C_{out}]$ & $[0,C_{in}]$ \\
        \textbf{out} & $[C_{out},0]$ & $[C_{out},C_{out}]$ & $[C_{out},C_{in}]$ \\
        \textbf{in} & $[C_{in},0]$ & $[C_{in},C_{out}]$ & $[C_{in},C_{in}]$ \\
    \end{tabular}
    \vspace{0.1cm}
    \caption{Payoff matrix when both players are susceptible}
    \label{tab:mask_game_payoffBOTH}
    
    \begin{tabular}{c|ccc}
        & \textbf{no} & \textbf{out} & \textbf{in}\\
        \hline
        \textbf{no} & $[C_i,C_i]$ & $[0,C_{out}+C_i]$ & $[C_i,C_{in}+C_i]$ \\
        \textbf{out} & $[C_{out}+C_i,C_i]$ & $[C_{out},C_{out}+C_i]$ & $[C_{out}+C_i,C_{in}+C_i]$ \\
        \textbf{in} & $[C_{in},C_i]$ & $[C_{in},C_{out}+C_i]$ & $[C_{in},C_{in}+C_i]$ \\
    \end{tabular}
    \vspace{0.1cm}
    \caption{Payoff matrix when exactly one player is susceptible}
    \label{tab:mask_game_payoffONE}
\end{table}

\begin{theorem}
\label{th:1}
When perfect knowledge is available about the states of the players, then if both players are of the same type, both the pure strategy Nash Equilibrium and the Social Optimum of the Mask Game are (\textbf{no}, \textbf{no}); while if exactly  one is susceptible (e.g., player 1) then the NE is (\textbf{in}, \textbf{no}) and the SO is (\textbf{no}, \textbf{out}).
\end{theorem}

In social optimum, susceptible players would benefit, through a positive externality, from an action that would impose a cost on infected players; therefore it is not a likely outcome. In fact, such a setting is common in man-made distributed systems, especially in the context of cybersecurity. A well-fitting parallel is defense against Distributed Denial of Service Attacks (DDoS) attacks~\cite{khouzani2013}: although it would be much more efficient to filter malicious traffic at the source (i.e., \textbf{out}), Internet Service Providers rather filter at the target (i.e., \textbf{in}) owing to a rational fear of free-riding by others. 

\subsection{Bayesian Game}

Since in the basic game no player plays \textbf{out}, we simplify the choice of the players to either \textbf{use} a mask or \textbf{no} (hence, we note the cost of a mask with $C_{use}$). To represent the situation more realistically, we introduce ambiguity about the status of the players: we denote the probability of being infected as $\rho$. We know from the basic game that if both players are infected (with probability $\rho^2$) or susceptible (with probability $(1-\rho)^2$) they play (\textbf{no,no}), while if only one of them is infected (with probability $2\cdot\rho\cdot(1-\rho)$) the infected player plays \textbf{no}, while the susceptible plays \textbf{use}. Hence, the players play \textbf{no} with probability $1-(\rho\cdot(1-\rho))$. In the following we analyse the case when these states are unknown. to the players. 

\begin{definition}
    The Bayesian Mask Game is a tuple $\langle\mathcal{N},\Sigma,\mathcal{U}\rangle$, where the set of players is $\mathcal{N}=\{1,2\}$ and their actions are $\Sigma=\{\textbf{no}, \textbf{use}\}$. The utility functions $\mathcal{U}=\{u_1,u_2\}$ are presented as a cost matrix in Table \ref{tab:util_bayes} where an additional term $+\rho C_i$ is missing element-wise for a clearer presentation.
\end{definition}

\begin{table}[t]
    \centering
    \begin{tabular}{c|cc}
        & \textbf{no} & \textbf{use}\\
        \hline
        \textbf{no} & $[\rho\cdot(1-\rho)\cdot C_i,\rho\cdot(1-\rho)\cdot C_i]$ & $[0,C_{use}]$ \\
        \textbf{use} & $[C_{use},0]$ & $[C_{use},C_{use}]$ \\
    \end{tabular}
    \vspace{0.1cm}
    \caption{The Payoff matrices of the Bayesian Mask Game}
\label{tab:util_bayes}
\end{table}

\begin{theorem}
\label{th:2}
When imperfect knowledge is available about the states of the players, then both the pure strategy Nash Equilibria and the Social Optima of the Mask Game are (\textbf{use}, \textbf{no}) and (\textbf{no}, \textbf{use}), respectively.
\end{theorem}

The Social Optimum evidently introduces unfairness, hence with adequate mechanism design  (\textbf{use}, \textbf{use}) could be reached (which is still better than (\textbf{no}, \textbf{no})) (see Section \ref{sec:mech}). Note, that within this paper we are focusing on the pure-strategy Nash Equilibrium (e.g., either \textbf{use} or \textbf{no}), however, it is possible that the game also has mixed-strategy Nash Equilibria (i.e., \textbf{use} with probability $\varphi$ and \textbf{no} with probability $(1-\varphi)$), which may lead to a utility increase. These randomized strategies could also be easily calculated; we leave this issue with the interested readers. We emphasize that our goal with these games are mainly demonstrative, therefore we deliberately do not provide a comprehensive game theoretical analysis. Moreover, in a sense, the uncertain nature of players' statuses have a similar effect on the outcome.

More realistic game models and their evaluation,  accounting for imperfect protection (Efficiency-Bayesian Game) and multiple participants, are described in Appendix~\ref{sec:effbay} and \ref{sec:multi}, respectively.

\section{The Distancing Game}
\label{sec:sd}

Another phenomenon most people has experienced during the current COVID-19 pandemic is social distancing. Here we introduce a simple Distancing Game to be played in sequence with the previously introduced Mask Game: once a player decided to meet up with friends via the Distancing Game, she can decide whether to wear a mask for the meeting by playing an appropriate version of the Mask Game. To improve readability, we summarize all corresponding parameters and variables in Table \ref{tab:parDIST}. 

We represent the cost of going out with $\rho\cdot m\cdot L$, i.e., the probability of getting infected (aka the infection rate) multiplied with the mortality rate of the disease and with the player's evaluation about her own life.\footnote{This is an optimistic approximation, as besides dying, the infection could impose other tolls on a player.} Besides the risk of getting infected, going out and attending a meeting could benefit the player, denoted as $B$. Consequently, while in the Mask Game we minimized the costs, here we maximize the payoff. In parallel to the benefit of a meeting, there is also a cost for staying home or missing a meeting, denoted as $C$. Of course, there are other alternative ways to capture the benefits and the cost of social distancing~\cite{thunstrom2020benefits}, but this simple utility function suffices for illustration purposes.

\begin{definition}
        The Distancing Game is a tuple $\langle\mathcal{N},\Sigma,\mathcal{U}\rangle$, where the set of players is $\mathcal{N}=\{1,2\}$, and their actions are $\Sigma=\{\textbf{go}, \textbf{stay}\}$. The utility functions $\mathcal{U}=\{u_1,u_2\}$ are presented as a payoff matrix in Table \ref{tab:util_dist}. 
\end{definition}

\begin{theorem}
\label{th:5}
A trivial Nash Equilibrium of the Distancing Game is (\textbf{stay}, \textbf{stay}), On the other hand (\textbf{go}, \textbf{go}) is also a NE if $\rho\cdot m\cdot L<B+C$. If this condition holds than (\textbf{go}, \textbf{go}) is also the Social Optimum, otherwise it is (\textbf{stay}, \textbf{stay}).
\end{theorem}

\begin{table}[t]
    \centering
    \begin{tabular}{c|l}
        Variable & Meaning \\
        \hline
        $C$ & Cost of staying home\\
        $B$ & Benefit of going out \\
        \hline
        $m$ & Mortality rate \\
        $L$ & Value of Life \\
        \hline
        $\rho$ & Probability of infection \\
        \hline
        $t$ & Time duration \\
    \end{tabular}
    \vspace{0.1cm}
    \caption{Parameters of the Distancing Games}
    \label{tab:parDIST}
    
    \centering
    \begin{tabular}{c|cc}
        & \textbf{go} & \textbf{stay}\\
        \hline
        \textbf{go} & $[B-\rho\cdot m\cdot L,B-\rho\cdot m\cdot L]$ & $[-\rho\cdot m\cdot L-C,-C]$ \\
        \textbf{stay} & $[-C, -\rho\cdot m\cdot L-C]$ & $[-C,-C]$ \\
    \end{tabular}
    \vspace{0.1cm}
    \caption{Payoff matrix of the Distancing Game}
\label{tab:util_dist}
\end{table}

\begin{example}
For instance, should a rational American citizen (e.g., Alice) meet Bob based on how much they value their lives? We estimate\footnote{Data from \url{https://www.worldometers.info/coronavirus/} (accessed 30th April, 2021)} $m=0.0225$, because $0.021\approx\frac{\#\{\text{deceased}\}}{\#\{\text{all cases}\}}<m<\frac{\#\{\text{deceased}\}}{\#\{\text{closed cases}\}}\approx0.024$ and $\rho=0.0025$ as in the previous examples.

Using these values, Alice should go out only if she values her life less than $17777(=\frac1{0.0225\cdot0.0025})$ times the sum of the benefit of the meeting and the loss of missing out. According to~\cite{trottenberg2013guidance}, the value of a statistical life in the US was 9.2 million USD in 2013, which is equivalent to 11.7 million USD in 2021 (with $0.3\%$ interest rate). This means that Alice should only meet someone if the benefit of the meeting plus the cost of missing it would amount to more than $658$ USD ($=\frac{11.7M}{17777}$).
\end{example}

We defined this game as symmetric, however, one can easily adapt the analysis to an asymmetric payoff structure (i.e., $B_1$, $B_2$, $C_1$, and $C_2$ instead of $B$ and $C$). Consider the example, where player 1 is selling a car because she urgently needs money, then $B_1=100$ and $C_1=90$ are feasible choices; while, for an average buyer, these parameters could be much lower, e.g., $B_2=50$ and $C_2=10$.

\subsection{Extended Distancing Game}
\label{sec:esd}

One way to improve the above model is by introducing meeting duration.\footnote{We capture the duration of the meeting equivalently as meeting size would be modelled, hence all our arguments about meeting duration could easily be adapted to optimal meeting size. } Leaving our disinfected home during a pandemic is risky, and this risk grows with the time. In the original model, we captured the infection probability with $\rho=1-(1-\rho)$. This ratio increases to $1-(1-\rho)^{t}$ for time $t$. We leave the interpretation of unit time to the reader. Moreover, the benefit of attending a meeting should depend on this new parameter, as well as the cost of isolation. For instance, staying home for a longer period might cause anxiety, which could get worse over time (i.e., increasing the cost)~\cite{venkatesh2020social}; on the other hand, spending longer quality time with someone could significantly boost the experience (i.e., increase the benefit). 

\begin{definition}
        The Extended Distancing Game is a tuple $\langle\mathcal{N},\Sigma,\mathcal{U}\rangle$, where the set of players is $\mathcal{N}=\{1,2\}$ and their actions are $\Sigma=\{\textbf{go}, \textbf{stay}\}$. The utility functions $\mathcal{U}=\{u_1,u_2\}$ are presented in Equation (\ref{eq:util_dist_ext}). 
\end{definition}

\begin{equation}
    \label{eq:util_dist_ext}
    u(\textbf{go})=\left\{
    \begin{tabular}{ll}
        $B(t)-(1-(1-\rho)^t)\cdot m\cdot L$ & if other \textbf{go} \\
        $-(1-(1-\rho)^t)\cdot m\cdot L-C(t)$ & if other \textbf{stay}
    \end{tabular}
    \right.\hspace{.2cm}
    u(\textbf{stay})=-C(t)
\end{equation}

A direct consequence of this extension is that the structure of the Distancing Game remained unchanged, hence, the two games share the same NEs. 

\begin{corollary}
\label{th:6}
Similarly to the basic Distancing Game, the Extended Distancing Game have the same trivial NEs (\textbf{stay}, \textbf{stay}) and (\textbf{go}, \textbf{go}) if $(1-(1-\rho)^t)\cdot m\cdot L<B(t)+C(t)$. If this condition holds than (\textbf{go}, \textbf{go}) is also the Social Optimum, otherwise it is (\textbf{stay}, \textbf{stay}).
\end{corollary}

\begin{figure*}[t]
    \centering
    \includegraphics[width=0.5\textwidth]{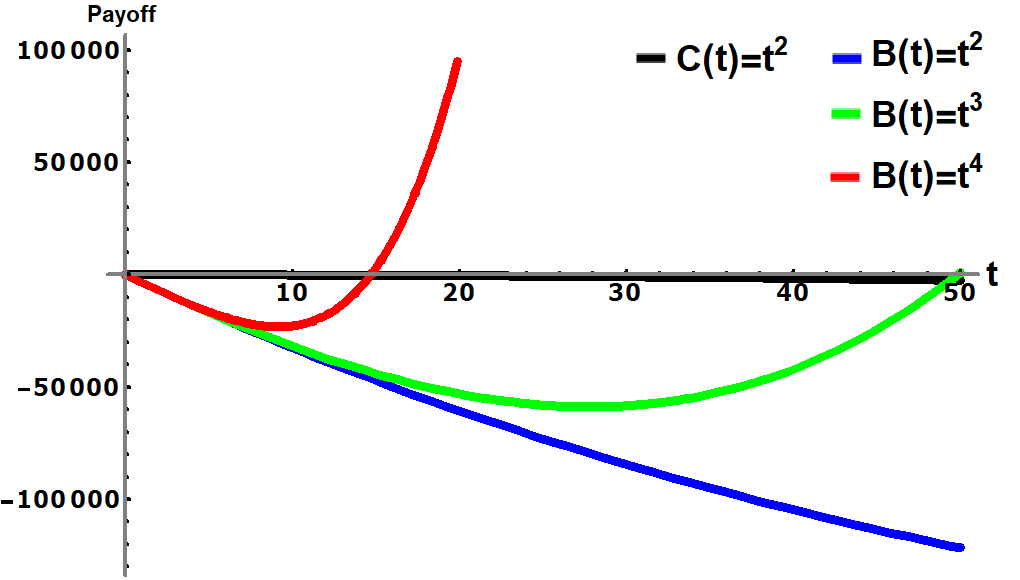}
    \caption{Illustrating the payoffs of the NEs (\textbf{stay}, \textbf{stay}) and (\textbf{go}, \textbf{go}) for the Extended Distancing Game with parameters $B=\{t^2,t^3,t^4\}$, $C(t)=t^2$, $m=0.0225$, $\rho=0.0025$, and $L=11,700,000$}
    \label{fig:dist_ext}
\end{figure*}

\begin{example}
In Figure \ref{fig:dist_ext}, we illustrate the payoffs for several polynomial benefit functions (e.g., $B(t)=\{t^2, t^3, t^4\}$) and a cost function $C(t)=t^2$. The rest of the parameters are defined as before, i.e., $\rho=0.0025$, $m=0.0225$, and $L=11,700,000$. It is visible that the change of the cost is insignificant compared to the benefit, consequently the illustration and the reasoning would be similar if $C$ would be linear or constant. It is visible that for small $t$ (\textbf{stay}, \textbf{stay}) is the SO as the utility is higher. On the other hand, as $t$ grows (\textbf{go}, \textbf{go}) becomes the SO. Another clearly visible take-away message is that the threshold of $t$ where the SO changes is lower as the benefit is higher. 
\end{example}

Note that not  even the extended distancing model captures behavioral characteristics of human decision making. For example, some people may live in a constant panic resulting in extreme risk aversion (i.e., staying at home even if it is irrational), while others may exhibit the ''it won't happen to me`` bias~\cite{erev2020complacency} resulting in (objectively) extreme risk taking (i.e., acting as if there was no pandemic).

\section{Vaccination Game}
\label{sec:vacc}

The most recent virus spreading prevention mechanism against the COVID-19 is vaccination. Since researching and developing a vaccine takes time, it could not be utilized as rapidly as the rest of the techniques detailed in this work (e.g., masks and social distancing). On the other hand, this protection mechanism is considered to be the most efficient and has proven its strength several times in the past~\cite{plotkin2005vaccines}. Concerning the rapidly developed COVID-19 vaccines, most governments and international organizations agree that all vaccines are safe to use and protect (to an extent) against COVID-19 for the general population. Yet, there are various aspects in which these vaccines differ, so individuals could have preferences.

\begin{table}[b]
    \centering
    \footnotesize
    \begin{tabular}{c|cccccc}
        Vaccine & Technology & Availability & Side-Effect & Efficiency & Duration & Usability \\
        \hline
        $\alpha$ & old & now & no & low &long & limited \\
        $\beta$ & new & soon & maybe & high & short & wide \\
    \end{tabular}
    \vspace{0.1cm}
    \caption{The two vaccines and their properties }
    \label{tab:vaccines}
\end{table}

Here, besides a few games, we also introduce several optimization models, where---in contrast to multiplayer games---individual's utility does not depend on other players' actions. The decision we model originates from the choice among multiple specific vaccines. Rather than focusing on whether to be vaccinated or not, as several previous works~\cite{bhattacharyya2011wait,bauch2004vaccination,van2008self} did, we compare two hypothetical vaccines, differing along 6 different dimensions as summarized in Table \ref{tab:vaccines}. The justification behind such a decision model is the real-world scenario prevalent in countries of Central and Eastern Europe, where as many as 8 different vaccines\footnote{\url{https://covid19.trackvaccines.org/country/hungary/}} were available at certain points in time. \textit{Technology} refers to the working mechanism of the vaccine (e.g., using dead/weakened virus, mRNS, etc)~\cite{le2020covid}. 
\textit{Availability} means the point in time when the vaccines are at the actual disposal of individual decision makers. It is reasonably expected that vaccines based on traditional technologies could be mass manufactured and transported with ease, while vaccines based on new technologies could be delayed for many reasons~\cite{khamsi2020if}. 
A similar difference corresponds to the potential \textit{side-effect}: vaccines based on older technologies were utilized in the past around the globe, hence the rare side-effects are either known or non-existing. On the other hand, side-effects concerning modern vaccines are only based on tests with a limited number of participants~\cite{ostergaard2021thromboembolism}. 
The \textit{efficiency} and \textit{duration} of the vaccines (e.g., the probability of mitigating the severe consequences of an infection and the length of the response of the body triggered by the vaccine, respectively) also differ, favouring the newer technology~\cite{wei2021impact}. 
Finally, the \textit{usability} of a vaccine refers to the portion of individuals who could/should get it, e.g., there are vaccines which were associated with severe side effects which effects various demographic groups differently \cite{mahase2021astrazeneca}. 
These differences between the two vaccines considered by the individuals are formalized in Table \ref{tab:parVACC} with the corresponding cost and benefit variables.

\begin{table}[b]
    \centering
    \begin{tabular}{l|cc}
        Vaccine & $\alpha$ & $\beta$ \\
        \hline
        Protection Efficiency & $e_\alpha$ & $e_\beta$ \\
        Effect Duration (time) & $d_\alpha$ & $d_\beta$ \\
        Availability (from time) & $0$ & $t_0$ \\
        Side-effect Probability & $0$ & $\epsilon$ \\
        Benefit of being vaccinated & $B_\alpha$ & $B_\beta$ \\
    \end{tabular}
    \vspace{0.1cm}
    \caption{Vaccine specific variables}
    \label{tab:vacc_ab}
\end{table}

\begin{table}[b]
        \centering
        \begin{tabular}{c|l}
        Variable & Meaning \\
        \hline
        $C_i$ & Cost of being infected\\
        $C_s$ & Cost of the side-effect\\
        $p$ & Vaccine preference\\
        \end{tabular}
        \caption{Costs \& Benefits of the Vaccination Game}
        \label{tab:util_par}
\end{table}

\begin{figure}[t]
     \centering
     \begin{subfigure}{0.3\textwidth}
         \centering
        \includegraphics[width=4.25cm]{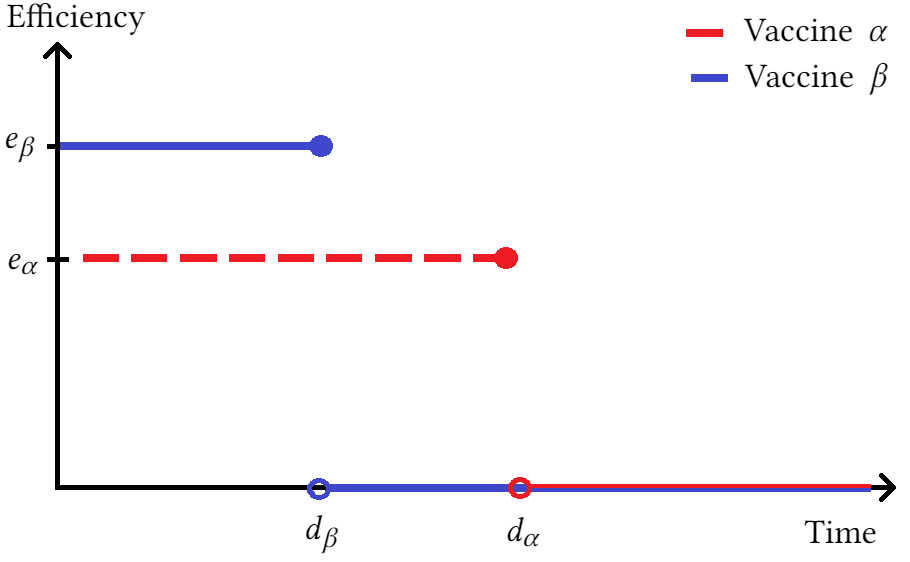}
        \caption{Duration and efficiency of the vaccines }
        \label{fig:vacc1}
     \end{subfigure}
     \begin{subfigure}{0.3\textwidth}
         \centering
        \includegraphics[width=4.25cm]{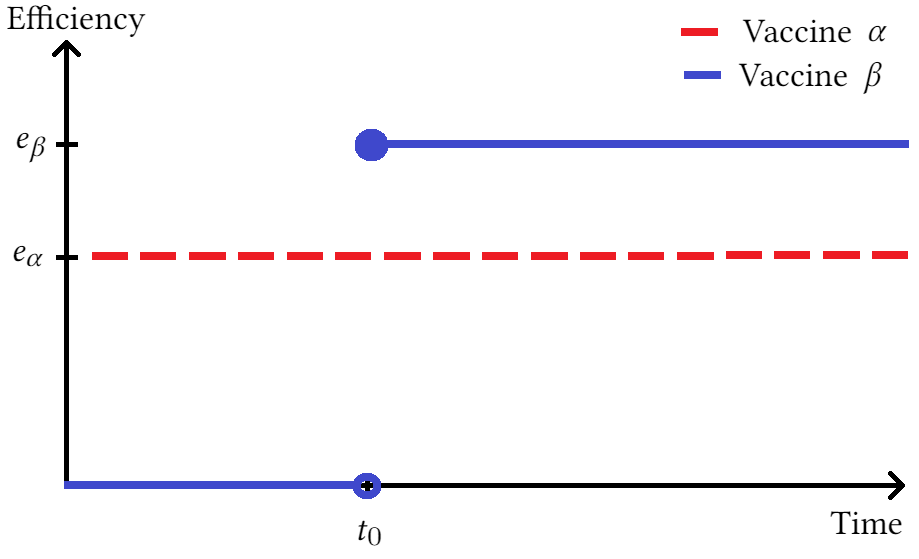}
        \caption{Availability and efficiency of the vaccines }
        \label{fig:vacc2}
     \end{subfigure}
     \begin{subfigure}{0.3\textwidth}
         \centering
        \includegraphics[width=4.25cm]{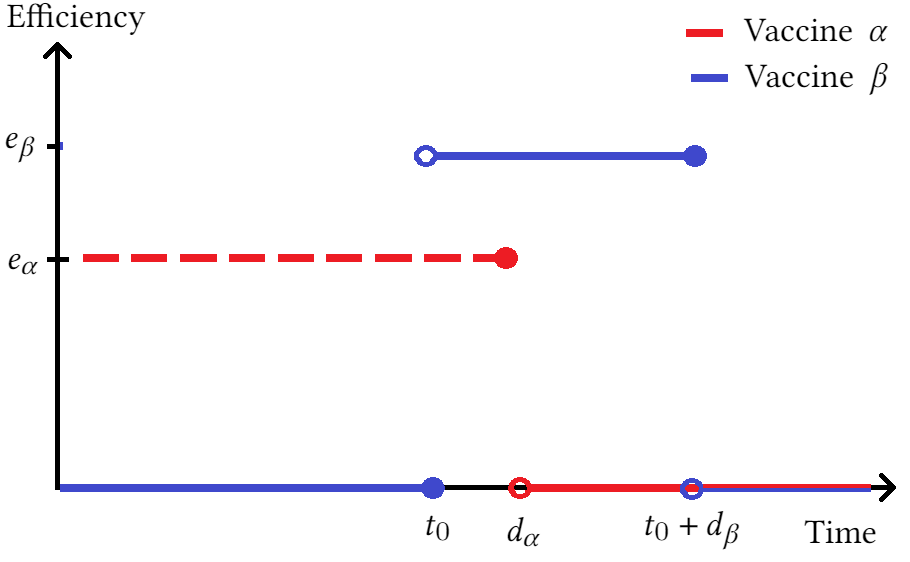}
        \caption{Availability, duration, and efficiency of the vaccines }
        \label{fig:vacc3}
     \end{subfigure}
    \caption{Illustration of vaccine properties}
    \label{fig:vacc}
\end{figure}

\subsection{Individual Optimization Models}

In the following optimization models we select 2-3 of the dimensions above, and present the utility/objective function for which the individuals optimize by selecting the vaccine with the higher payoff. We do not provide formal theorems and proofs as the results are trivial corollaries of the exact definitions. 

\paragraph{Duration-Efficiency Decision. } 
As defined in Table \ref{tab:vacc_ab} we assume vaccine $\alpha$ provides protection for  duration $d_\alpha$ with protection level $e_\alpha$. On the other hand, we assume that Vaccine $\beta$ protects for a shorter duration $d_\beta$ but with a stronger protection level $e_\beta$. This is also illustrated in Figure \ref{fig:vacc1}.

\begin{definition}
        The Duration-Efficiency decision problem is a tuple $\langle\Sigma,\mathcal{U}\rangle$, where the actions are $\Sigma=\{\alpha,\beta\}$ and the corresponding utility functions $\mathcal{U}=\{U(\alpha),U(\beta)\}$ are presented in Equation (\ref{eq:vacc1}):
\end{definition}

\begin{equation}
    \label{eq:vacc1}
    U(\alpha)=\int_0^{d_\alpha} e_\alpha\di t=e_\alpha\cdot d_\alpha \hspace{2cm}
    U(\beta)=\int_0^{d_\beta} e_\beta\di t=e_\beta\cdot d_\beta
\end{equation}

It is clear that the optimal decision depends on the exact values of $e_\alpha$, $e_\beta$, $d_\alpha$, and $d_\beta$: if $e_\alpha\cdot d_\alpha>e_\beta\cdot d_\beta$ then Vaccine $\alpha$ is the optimal choice, otherwise it is Vaccine $\beta$. For instance if we set $e_\alpha=0.76$, $e_\beta=0.95$, $d_\alpha=49$ and $d_\beta=35$ then $U(\alpha)\approx37>33\approx U(\beta)$.\footnote{These exact values are illustrative and do not correspond to any existing vaccines. We fabricated these numbers using the following sources:\\
\tiny
\url{https://qubit.hu/2021/05/19/az-astrazeneca-legolcsobb-a-sinopharm-a-legdragabb-a-vakcina-vilagpiacon}\\
\url{https://qubit.hu/2021/04/26/a-pfizer-antitestes-vedelme-erosebb-de-hamarabb-elhalvanyul-mint-az-astrazenecae}\\
\url{https://qubit.hu/2021/03/31/pfizer-szputnyik-sinopharm-mennyire-vedenek-a-vakcinak-a-brit-es-a-del-afrikai-varians-ellen}.}

\paragraph{Availability-Efficiency Decision. }
Following Table \ref{tab:vacc_ab} we assume Vaccine $\alpha$ is available now (i.e., at $t=0$), but it only provides protection level  $e_\alpha$. On the other hand, Vaccine $\beta$ will only become available at $t_0$, but with a stronger protection level $e_\beta$. This is also illustrated in Figure \ref{fig:vacc2}. Even without taking the duration into account, we have to introduce time-based discounting for the utility via the factor  $\delta$, as generally treated in the economics literature~\cite{frederick2002time}. 

\begin{definition}
        The Availability-Efficiency decision problem is a tuple $\langle\Sigma,\mathcal{U}\rangle$, where the actions are $\Sigma=\{\alpha,\beta\}$, and the corresponding utility functions $\mathcal{U}=\{U(\alpha),U(\beta)\}$ are presented in Equation (\ref{eq:vacc2}):
\end{definition}

\begin{equation}
    \label{eq:vacc2}
    U(\alpha)=\int_0^\infty e_\alpha\cdot\delta^t\di t=\frac{-e_\alpha}{\log \delta} \hspace{2cm}
    U(\beta)=\int_{t_0}^\infty e_\beta\cdot\delta^t\di t=\frac{-e_\beta}{\log \delta}\delta^{t_0}
\end{equation}

Again, the optimal decision trivially depends on the exact values of $e_\alpha$, $e_\beta$, $t_0$, and $\delta$: if $e_\alpha<e_\beta\cdot\delta^{t_0}$ then Vaccine $\alpha$ is the optimal choice, otherwise it is Vaccine $\beta$. For instance, with $e_\alpha=0.76$, $e_\beta=0.95$, $t_0=28$, and $\delta=0.999$, the utilities are $U(\alpha)\approx1749$ and $U(\beta)\approx2126$, respectively. 

\paragraph{Duration-Efficiency-Availability Decision. }
It is possible to combine the previous two decision models as illustrated in Figure \ref{fig:vacc3}.

\begin{definition}
        The Duration-Efficiency-Availability decision problem is a tuple $\langle\Sigma,\mathcal{U}\rangle$, where the actions are $\Sigma=\{\alpha,\beta\}$, and the corresponding utility functions $\mathcal{U}=\{U(\alpha),U(\beta)\}$ are presented in Equation (\ref{eq:vacc3}):
\end{definition}

\begin{equation}
    \label{eq:vacc3}
    U(\alpha)=\int_0^{d_\alpha} e_\alpha\cdot\delta^t\di t=(\delta^{d_\alpha}-1)\cdot\frac{e_\alpha}{\log(\delta)} \hspace{.2cm}
    U(\beta)=\int_{t_0}^{t_0+d_\beta} e_\beta\cdot\delta^t\di t=(\delta^{d_\beta}-1)\cdot\delta^{t_0}\cdot\frac{e_\beta}{\log(\delta)}
\end{equation}

The optimal decision depends on the exact values of $e_\alpha$, $e_\beta$, $d_\alpha$, $d_\beta$, $t_0$, and $\delta$: if $\frac{e_\alpha}{e_\beta}\cdot\delta^{t_0}>\frac{\delta^{d_\alpha}-1}{\delta^{d_\beta}-1}$ then Vaccine $\alpha$ is the optimal choice, otherwise it is Vaccine $\beta$. For instance, with $e_\alpha=0.76$, $e_\beta=0.95$, $d_\alpha=49$, $d_\beta=35$, $t_0=28$, and $\delta=0.999$, the utilities are $U(\alpha)\approx84$ and $U(\beta)\approx73$, respectively. 

It is clear that other factors also impact the optimal decision in a real-world setting. One of these are side-effects, described in Appendix~\ref{sec:side}.

\subsection{Availability Game}

Instead of combining more-and-more dimensions to formalize various utility / objective functions for the optimization models above, we rather include the other players' actions into the individual utility, as well. In contrast to the binary decision of the optimization problems, here player $n$'s action (out of $N$ players) is to set the preference $p_n\in[0,1]$, where 0 corresponds to Vaccine $\alpha$, and 1 corresponds to Vaccine $\beta$. Consequently, the availability (i.e., vaccination time) also changes dynamically based on this strategy: instead of 0 and $t_0$, it is set to player $n$ as $t_n=p_n\cdot t_0$. For the sake of simplicity, we use the benefit variables directly, instead of the efficiency and duration, to express the benefit of being vaccinated with a specific vaccine.

\begin{definition}
         The Availability Game is a tuple $\langle\mathcal{N},\Sigma,\mathcal{U}\rangle$, where the set of players is $\mathcal{N}=\{1,\dots,N\}$, and their actions are $\Sigma=\{p_1,\dots,p_N\}$,m where $\forall$ $n: p_n\in[0,1]$. The utility functions $\mathcal{U}=\{u_1,\dots,,u_N\}$ are presented in Equation (\ref{eq:vacc_game1}) where the player's vaccination times $\{t_1, \dots, t_N\}$ are ordered such that $0=\hat{t}_0<\hat{t}_1\le\dots\le\hat{t}_{N}<\hat{t}_{N+1}=\infty$ and $t_n=\hat{t}_{n^\prime}$:
\end{definition}

\begin{equation}
    \label{eq:vacc_game1}
    U_n(p_1,\dots,p_N)=\int_{p_n\cdot t_0}^{\infty} (p_n\cdot B_\beta + (1-p_n)\cdot B_\alpha)\cdot \delta^t \di t - 
    \sum_{i=0}^{n^\prime-1}\int_{\hat{t}_i}^{\hat{t}_{i+1}}C_i\cdot\left(1-\frac{i}{N}\right)\cdot\delta^t \di t
\end{equation}

\begin{figure}[t]
    \centering
    \includegraphics[width=6cm]{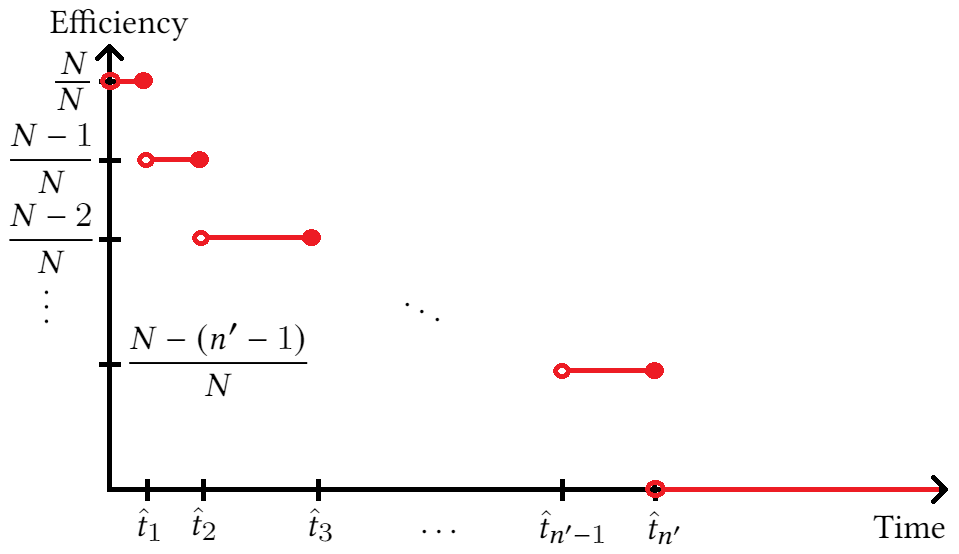}
    \caption{Illustration of the protection for the $n$ player Availability Game }
    \label{fig:inf}
\end{figure}

The benefit in Equation (\ref{eq:vacc_game1}) is constructed via a straightforward linear combination of the benefit of the vaccines. The cost is a little trickier though as the other player's actions do matter.\footnote{We ignore the effect of herd immunity, i.e., others' actions do not influence the benefit of a single player.} We assume that the vaccines decrease the spreading equally (hence the $1-\frac{i}{N}$ part), and provide perfect protection to the vaccinated (hence the summation goes only till $n^\prime-1$). This is also illustrated in Figure \ref{fig:inf}. Note that these assumptions can be easily relaxed by introducing efficiency/duration and other properties of the vaccines to the model. Moreover, the model can be extended with side-effects as well, but for the sake of simplicity we leave these for future work. 

\begin{lemma}
\label{th:7}
The utility function defined in Equation (\ref{eq:vacc_game1}) is equivalent to Equation (\ref{eq:av_game1}).
\end{lemma}

\begin{equation}
    \label{eq:av_game1}
    U_n(p_1,\dots,p_N)=\frac{\delta^{p_n\cdot t_0}\cdot(p_n\cdot B_\beta + (1-p_n)\cdot B_\alpha)}{-\log(\delta)}-\frac{C_i\cdot(N-\delta^{\hat{t}_1}-\dots-\delta^{\hat{t}_{n^\prime}})}{N\cdot\log(\delta)}
\end{equation}

\begin{theorem}
\label{th:8}
The utility in the Availability Game when the player's actions are symmetric (i.e., $\forall$ $n\in[1,N]: p_n=p$) is shown in Equation (\ref{eq:av_game2}). In this case the NE is one of the following:
\begin{itemize}
    \item $p^*=\frac{C_i + B_\alpha\cdot N}{(-B_\beta + B_\alpha)\cdot N} - \frac1{t_0\cdot \log(\delta)}$, if $0<p^*<1$.
    \item $p=0$ if $p^*\not\in[0,1]$ and $B_\beta\cdot\delta^{t_0}-B_\alpha< \frac{C_i}{N}\cdot(1-\delta^{t_0})$.
    \item $p=1$ if $p^*\not\in[0,1]$ and $B_\beta\cdot\delta^{t_0}-B_\alpha> \frac{C_i}{N}\cdot(1-\delta^{t_0})$.
\end{itemize}
\end{theorem}

\begin{equation}
    \label{eq:av_game2}
    U_n(p)=\frac{\delta^{p\cdot t_0}\cdot(p\cdot B_\beta + (1-p)\cdot B_\alpha)}{-\log(\delta)}-\frac{C_i\cdot(N-\delta^{p\cdot t_0})}{N\cdot\log(\delta)}
\end{equation}

\begin{figure}[t]
    \centering
    \includegraphics[width=12cm]{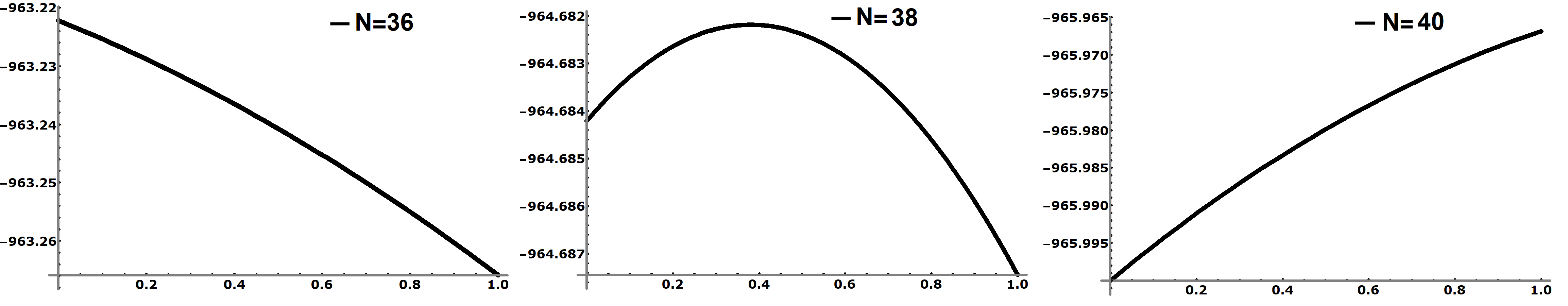}
    \caption{Illustration of the importance of the amount of players when $B_\alpha=9$, $B_\beta=10$, $C_i=1000$, $t_0=28$, and $\delta=0.999$. From left to right the utility is shown for $N=\{36,38,40\}$ }
    \label{fig:ava}
\end{figure}

As the theorem states, the NE depends on the exact values of $B_\alpha$, $B_\beta$, $C_i$, $t_0$, and $\delta$. For instance, with $B_\alpha=9$, $B_\beta=10$, $C_i=1000$, $t_0=28$, and $\delta=0.999$, the utility is shown in Figure \ref{fig:ava} for $N=\{36,38,40\}$, respectively. The three sub-figures represent the three cases of the theorem. It is visible that with these parameters, with a low number of participants ($N<36$), players prefer to vaccinate later; while with a high number of participants ($N>40$), the preferred choice is early vaccination. 

\section{Pandemic Mechanism Design}
\label{sec:mech}

The three games described above model only parts of the bigger picture.


\subsection{The government as mechanism designer}

 We refer to the collection (and interplay) of measures implemented by a specific government fighting the epidemic in their respective country as \emph{mechanism}. Consequently, decisions made with regard to this mechanism constitutes \emph{mechanism design}~\cite{mas1995microeconomic}. In its broader interpretation, mechanism design theory seeks to study mechanisms achieving a particular preferred outcome. Desirable outcomes are usually optimal either from a social aspect or maximizing a different objective function of the designer.

In the context of the coronavirus pandemic, the immediate response mechanism is composed of, e.g., wearing a mask, social distancing, testing and contact tracing, among others, followed by vaccination. Note that this is not an exhaustive list: financial aid, creating extra jobs to accommodate people who have just lost their jobs, declaring a national emergency and many other conceptual vessels can be utilized as sub-mechanisms by the mechanism designer, i.e., usually, the government; we do not discuss all of these in detail due to the lack of space. Instead, we shed light on how government policy can affect the sub-mechanisms, how sub-mechanisms can affect each other and, finally, the outcome of the  mechanism itself. We illustrate the importance of mechanism design applying different policies to our three games, and adding testing and contact tracing to the mix.

\subsection{Policy impact on sub-mechanisms and the final mechanism}

Here we analyze the impact of specific policies on data transparency, mask wearing, social distancing, testing and contact tracing, and vaccination. 

\paragraph{Data quality/transparency. }
It is well-known that inaccurate reporting of epidemic data can potentially decrease the efficacy of forecasting, and thus, response measures~\cite{hong2020estimation}. A less understood aspect of the data quality problem is the deliberate distortion of such reports. While not specific to handling the COVID-19 situation, a government's decision to be fully transparent or to partially conceal information from its citizens could have a profound impact on the success of pandemic response. It is fairly straightforward to see that if people make their individual decisions based on deliberately manipulated, coarse-grained or gappy data, the results will be sub-optimal and, potentially even more detrimental, unpredictable. If there is no unanimously trusted source of information available, people's beliefs will be heterogeneous, as if they were playing different games altogether. As a simple example, take the Distancing game in Section \ref{sec:sd}, where individuals will make their assessments whether to meet based on $\rho$, the probability of getting infected. If media reports on this parameter are altered or varying across different channels, people may a) meet up when it is not in their best interest, or b) stick to staying home even if it is no longer sensible. While the detrimental effect of data concealment seems rather indirect and hard to concretize, there exist quantitative reports aiming to shed light on such issues, e.g., on data concealment and COVID-19 mortality~\cite{toth21}.

\paragraph{Compulsory mask wearing and free masks. }If the government declares that wearing a simple mask is mandatory in public spaces (such as shops, mass transit, etc.), it can enforce an outcome (\textbf{out}, \textbf{out}) that is indeed socially better than the NE. The resulting strategy profile is still not SO, but it i) allocates costs equally among citizens; ii) works well under the uncertainty of one's health status; and iii) may decrease the first-order need for large-scale testing, which in turn reduces the response cost of the government. By distributing free masks, the government can reduce the effect of selfishness and, potentially, help citizens who cannot buy or afford masks owing to supply shortage or unemployment.

\paragraph{Limiting the amount of people gathering and total lock-down. }Within the Extended Distancing Game, the time parameter $t$ captures the duration of a meeting. This could have another interpretation as well, as meeting size could be captured the same way as time. Consequently, if the government imposes an upper limit $T$ for the size of congregations, this will put a strict upper bound on the ``optimal meeting size'' $t^*$, and the resulting group size will be $\min(T,t^*)$, instantiating a decreased benefit, and, therefore, promoting staying at home. On the other hand, if the chosen restrictive measure is a total lock-down, both the Distancing Game and the Mask Game are rendered moot, as people are not allowed to leave their households.

\begin{figure}[t]
    \centering
    \includegraphics[width=0.95\textwidth]{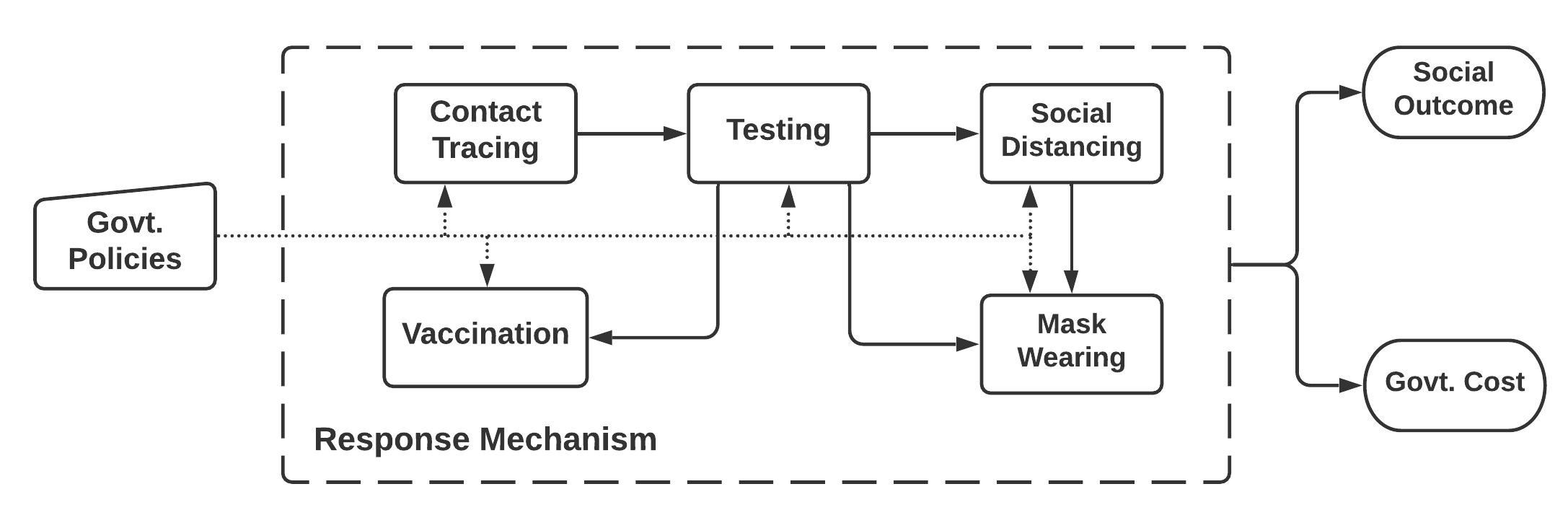}
    \caption{Pandemic response mechanism as influenced by government policy (dotted lines) and the interplay of sub-mechanisms (solid lines)}
    \label{fig:bigpicture}
\end{figure}

\paragraph{Testing and contact tracing. }It is clear that the Distancing and the Mask Games are not played in isolation: people deciding to meet up invoke the decision situation on mask wearing. On the other hand, so far we have largely ignored three other widespread pandemic response measures: testing, contact tracing, and vaccination.

With appropriately designed and administered coronavirus tests, medical personnel can determine two distinct features of the tested individual: i) whether she is actively infected spreading the virus and ii) whether she has already had the virus, even if there were no or weak symptoms. (Note that detecting these two features require different types of tests, able to show the presence of either the virus RNA or specific antibodies, respectively.) In general, testing enables both the tested person and the authorities to make more informed decisions. Putting this into the context of our games, testing i) reduces the uncertainty in Bayesian decision making, ii) enables the government to impose mandatory quarantine thereby removing infected players, and iii) identifies individuals who are temporarily immune, and thus, can be vaccinated at a later stage without imposing greater risk on them.

Even more impactful, mandatory testing (as in Wuhan\footnote{New York Times. \url{https://www.nytimes.com/2020/05/26/world/asia/coronavirus-wuhan-tests.html}}) completely eliminates the Bayesian aspect, essentially rendering the situation to a full information game: 
it serves as an exogenous ``health oracle'' imposing no monetary cost on the players. To sum it up, the testing sub-mechanism outputs results that serve as inputs to the Distancing, Mask and Vaccination Games.

Naturally, a ``health oracle'' does not exist: someone has to bear the costs of testing. From the government's perspective, mandatory mass testing is extremely expensive\footnote{But not without precedence, e.g., in Slovakia (\url{https://edition.cnn.com/world/live-news/coronavirus-pandemic-10-18-20-intl/h_beb93495fe9b83701023eafd5f28e39d})}. (Similarly, from the concerned individual's perspective, a single test could be unaffordable.) Contact tracing, whether traditional or mobile app-based, serves as an important input sub-mechanism to testing~\cite{ferretti2020quantifying}. It identifies the individuals who are \emph{likely} affected based on spatial proximity, and inform both them and the authorities about this fact. In game-theoretic terms, for such players, the benefit of testing outweigh the cost (per capita) with high probability. From the mechanism designer's point of view, contact tracing reduces the overall testing cost by enabling \emph{targeted testing}, potentially by orders of magnitude, without sacrificing proper control of the pandemic. Another potential cost of contact tracing for individuals could be the loss of privacy. Note that mobile OS manufacturers are working on integrating privacy-preserving contact tracing into their platform to eliminate adoption costs for installing an app\footnote{Apple. \url{https://covid19.apple.com/contacttracing}}.

\paragraph{Vaccination. }By far, vaccination policy is the most complicated and scrutinized among all sub-mechanisms, owing to its direct relation to control over one's own body, a pillar of human rights.

The availability of multiple, high efficacy vaccines enables governments to contain and suppress the pandemic. It is clear that, even if herd immunity is never reached, the more people are vaccinated, the less problem COVID-19 will cause in the near future. As mandatory vaccination is not feasible even in semi-democracies, the design of an efficient carrot-and-stick system is sensible. Therefore, countries have started to introduce vaccination passports~\cite{phelan2020covid}, which give to its holders benefits over their non-vaccinated countrymen, such as attending indoor venues, mass events like concerts or football matches, and traveling internationally without continuous testing. Sensibility notwithstanding, even the vaccine passport concept is under heavy legal and ethical scrutiny (not to mention compulsory vaccination).  

As vaccines have so far been a scarce resource, government decision on which vaccines to purchase in what quantities can be crucial. Exacerbated by incomplete trial documentation, the lack of trust between countries, being in different stages of the pandemic, and having greatly varying financial and healthcare means available, national governments have followed different strategies. In a country, where the pandemic is fairly well-contained with mild restrictive measures, playing it safe makes perfect sense\footnote{\url{https://www.fhi.no/en/id/vaccines/coronavirus-immunisation-programme/}}. In such a case, the Availability Game might take a special form, where there is only 1 type of vaccine available (or multiple, but equivalent vaccines), and the vaccination period is elongated. However, it is in the best interest of a country with high mortality and collapsing healthcare to grab any available, perhaps under-documented or lower efficacy vaccine in significant quantities. In the latter scenario, there might be 5-6 different types of vaccines in a national vaccination program\footnote{\url{https://abouthungary.hu/news-in-brief/coronavirus-heres-the-latest}}.

Adding to the set of available vaccines, the proposed order of vaccination is another important control lever. Most implemented policies agree on prioritizing medical staff and emergency first responders, but can differ on prioritizing the elderly (demographic segment with the highest risk of death/severe symptoms) or the actively working people (segment with the highest risk of transmission)~\cite{wang2021assessing}. Combining this aspect with the individual preference for a certain type of vaccine, the part of the population that does not want to be vaccinated, and the uncertainty of to which extent vaccines prevent transmission, realistically, the mechanism designer can only aim for an approximately optimal policy design. Adding to this, the right policy for relaxing restrictive measures as the vaccination progresses constitutes an issue of its own~\cite{wang2021assessing}, and has an effect on all the sub-mechanisms and games mentioned above.

\paragraph{The big picture. }As far as pandemic response goes, the mechanism designer has the power to design and parametrize the games that citizens are playing, taking into account that sub-mechanisms affect each other. After games have been played and outcomes have been determined, the cost for the mechanism designer itself are realized (see Figure~\ref{fig:bigpicture}). This cost function is very complex incorporating factors from ICU beds through civil unrest and affected future election results to a drop in GDP over multiple time scales~\cite{mcdonald2008macroeconomic}. Therefore, governments have to carefully balance the---very directly interpreted---social optimum and their own costs; this indeed requires a mechanism design mindset.

\section{Conclusion}
\label{sec:con}

In this paper we have made a case for treating pandemic response as a mechanism design problem. Through simple games modeling interacting selfish individuals we have shown that it is necessary to take individual incentives into account during a pandemic. First, we have shown how individual incentives impact mask wearing, both when players have perfect knowledge on their own and co-players' health status, and when there is uncertainty involved. Second, we have shown how individual decisions (and, therefore, social impact) concerning social distancing depend on the perceived benefits of meeting up and the cost of missing out. Third, we have illustrated how individuals could optimize when selecting between two hypothetical vaccines taking into account availability, efficacy and duration of immunity. Moreover, we derived how the attractiveness of early or late vaccination depends on vaccine characteristics and the number of players. 

We have also demonstrated that specific government policies significantly influence the outcome of these games, and how different response measures (sub-mechanisms) are interdependent. As an example we have discussed how contact tracing enables targeted testing which in turn reduces the uncertainty from individual decision making regarding social distancing and wearing masks. Furthermore, we have discussed the notoriously complex nature of the vaccination policy; designing such in an even approximately optimal way has to take into account medical, behavioral, economic and legal factors. We have also argued that sharing high quality and truthful pandemic data with the public promotes better individual decision making, and thus, more efficient handling of the pandemic. Governments have significantly more power than traditional mechanism designers in distributed systems; therefore, it is even more crucial for them to carefully study the tradeoff between social good and the cost of the designer when implementing their pandemic response mechanism.

\paragraph{Limitations and future work. }
The work presented here has several limitations from a policy-making standpoint. First, although the mechanism designer can directly influence the payoff functions and thus the outcome of the games presented (e.g., by imposing fines on non-compliant citizens or giving benefits to the vaccinated), and the factors currently used in the payoffs ---without doubt---do play a part in individual decisions making, the utility functions themselves are---of course---simplified: behavioral decision-making aspects are out of scope for this paper. Second, at this level of abstraction, the games and their respective designs cannot form a practical guidebook for governments. In fact, complex simulation studies and the analysis of already existing real historical data have to be undertaken in order to make real-world decisions affecting human lives. Recall that the objective of this study is to \emph{illustrate the impact of individual decision making} on widespread pandemic measures, and \emph{advocate for a mechanism design mindset} for policy-makers.

We have barely scratched the surface of pandemic mechanism design. The models presented are simple and mostly used for demonstrative purposes. In turn, this gives us plenty of opportunity for future work. A potential avenue is extending our models to capture the temporal aspect, combine them with epidemic models as games played by many agents on social graphs, and parametrize them with real data from the ongoing pandemic (policy changes, mobility data, price fluctuations, etc.). Relaxing the rational decision-making aspect is another prominent direction: behavioral modeling with respect to obedience, other-regarding preferences and risk-taking could be incorporated into the games. Finally, a formal treatment of the mechanism design problem constitutes important future work, incorporating hierarchical designers (WHO, EU, nations, municipality, household), an elaborate cost model, and analyzing optimal policies for different time horizons. If done with care, these steps would help create an extensible mechanism design framework that can aid decision makers in pandemic response.

\section*{Acknowledgements}
Project no. 138903 has been implemented with the support provided by the Ministry of Innovation and Technology from the NRDI Fund, financed under the FK\_21 funding scheme.\\
This work was supported by the National Research, Development and Innovation Fund of Hungary in the frame of FIEK\_16-1-2016-0007 (Higher Education and Industrial Cooperation Center) project.

\bibliographystyle{plain}
\bibliography{ref.bib}

\appendix
\section{Appendix}

\subsection{Efficiency-Bayesian Mask Game}
\label{sec:effbay}

In the Mask Game we assumed \textbf{in} provides perfect protection from infected players, while \textbf{out} protects the other player fully. However, in real life, these strategies mitigate the infection by decreasing its probability only to some extent. For this reason, we define $a, b\in[0,1]$: $a$ measures the protection efficiency of the protection strategy, while $b$ captures the efficiency of eliminating the further spread of the disease. Consequently, $a$ and $b$ were set in the previous cases to $a_\textbf{out}=0$ (i.e., \textbf{out} has no effect on protecting the player), $a_\textbf{in}=1$ (i.e., \textbf{in} perfectly protects the player), and $b_\textbf{in}=1,b_\textbf{out}=0$ (i.e., \textbf{in} prevents further spreading, while \textbf{out} does not). We simplify the action space of the players as we did in the Bayesian Game: \textbf{in} and \textbf{out} is merged into \textbf{use}. Obviously, \textbf{no} corresponds to $a_{no}=b_{no}=0$. In the following we slightly abuse the notation with $a$ and $b$ representing $a_{use}$ and $b_{use}$, respectively. 

\begin{definition}
        The Efficiency-Bayesian Mask Game is a tuple $\langle\mathcal{N},\Sigma,\mathcal{U}\rangle$, where the set of players is $\mathcal{N}=\{1,2\}$ and their actions are $\Sigma=\{\textbf{no}, \textbf{use}\}$. The utility functions $\mathcal{U}=\{u_1,u_2\}$ are presented as a payoff matrix in Table \ref{tab:util_eff}, where an additional term $+\rho\cdot C_i$ is missing element-wise for a clearer presentation, and $\hat{C}_i=\rho\cdot(1-\rho)\cdot C_i$. 
\end{definition}

\begin{table}[b!]
    \centering
    \footnotesize
    \begin{tabular}{c|cc}
        & \textbf{no} & \textbf{use}\\
        \hline
        \textbf{no} & $[\hat{C}_i,\hat{C}_i]$ & $[(1-b)\cdot\hat{C}_i,(1-a)\cdot\hat{C}_i+C_{use}]$ \\
        \textbf{use} & $[(1-a)\cdot\hat{C}_i+C_{use},(1-b)\cdot\hat{C}_i]$ & $[C_{use}+(1-a)\cdot(1-b)\cdot\hat{C}_i,(1-a)\cdot(1-b)\cdot\hat{C}_{use}]$ \\
    \end{tabular}
    \vspace{0.1cm}
    \caption{Payoff matrix of the Efficiency-Bayesian Mask Game}
\label{tab:util_eff}
\end{table}

\begin{theorem}
\label{th:3}
When imperfect knowledge is available about the states of the players, and the mask's efficiency is known, then the pure strategy Nash Equilibrium of the Mask Game depends on the relation between the ratio $\frac{C_{use}}{C_i}$ and the interval $[a\cdot(1-b)\cdot\rho\cdot(1-\rho),a\cdot\rho\cdot(1-\rho)]$, specifically:
\begin{itemize}
    \item The NE is (\textbf{use}, \textbf{use}) if $\, \frac{C_{use}}{C_i} < a\cdot(1-b)\cdot\rho\cdot(1-\rho)$;
    \item The NEs are (\textbf{no}, \textbf{use}) and (\textbf{use}, \textbf{no}) if $\, a\cdot(1-b)\cdot\rho\cdot(1-\rho)< \frac{C_{use}}{C_i} < a\cdot\rho\cdot(1-\rho)$;
    \item The NE is (\textbf{no}, \textbf{no}) if $\, \frac{C_{use}}{C_i} > a\cdot\rho\cdot(1-\rho)$.
\end{itemize}
\end{theorem}

\begin{example}
We set $b=\frac23$, as surgical masks on the infectious person reduce cold \& flu viruses in aerosols around 70\% according to~\cite{milton2013influenza}. Parameter $a$ is much harder to measure. It should be $a\le b$ since any mask keeps the virus inside the players more efficiently than stopping the wearer from getting infected. For the sake of this example we set $a=\frac{b}{2}=\frac13$, but any other choice would be possible. We set $\rho\approx\frac{\#\{\text{active cases}\}}{\#\{\text{population}\}}\approx0.0025$ according to the the statistics of COVID-19 cases on 30th April, 2021\footnote{\url{https://www.worldometers.info/coronavirus/}}. With these values the interval is $[0.00055416,0.0016625]$, i.e., if the cost of being infected is more than the price of 1800 masks, then both players choose \textbf{use}, if it is more than 600 but less than 1800 then only one, and if it is less than 600 then neither one of them.
\end{example}

\subsection{Multiplayer Mask Game}
\label{sec:multi}

This game can be further extended by allowing $N$ players to participate in a meeting where they get in close contact with random $g$ players. In this extension we assume---without taking any spatial information into account---that all players meet with probability $\frac1N$. 

\begin{definition}
    The Multiplayer Mask Game is a tuple $\langle\mathcal{N},\Sigma,\mathcal{U}\rangle$, where the set of players is $\mathcal{N}=\{1,2,\dots,N\}$ and their actions are $\Sigma=\{\textbf{no}, \textbf{use}\}$. The utility (cost) functions $\mathcal{U}=\{u_n\}_{n=1}^N$ are presented in Equation (\ref{eq:multi_mask}) for both susceptible and infectious players,  where $k$ other players are infectious. 
\end{definition}

\begin{equation}
    \label{eq:multi_mask}
    u_n^{inf}=\left\{
    \begin{tabular}{ll}
        $C_i$ & if $n$ \textbf{no} \\
        $C_i+C_{use}$ & if $n$ \textbf{use}
    \end{tabular}
    \right.\hspace{.2cm}
    u_n^{sus}=\left\{
    \begin{tabular}{ll}
        $\left(1-\left(1-\frac{k}{N}\right)^g\right)\cdot C_i$ & if $n$ \textbf{no} \\
        $C_{use}$ & if $n$ \textbf{use}
    \end{tabular}
    \right.
\end{equation}

The cost corresponding to infectious players are trivial, while the cost of a susceptible player is either $C_{use}$ or the cost of infection multiplied with the probability of complement of none of the random $g$ players are infected. As a direct consequence of this utility function and the previous theorem the corresponding NE is trivial. 

\begin{corollary}
\label{th:4}
When perfect knowledge is available about the states of the players, then the best response of an infectious player is to play \textbf{no}. The susceptible player also plays \textbf{no} if $1-(1-\frac{k}{N})^g<\frac{C_{use}}{C_i}$. 
\end{corollary}

\begin{example}
The ratio $\frac{k}N$ is essentially the infection rate $\rho$ from the Bayesian Mask Game, hence we can define it as $\frac{k}N=\rho=0.0025$. Consequently, $C_i$ must be at least 400, 200, 100, and 50 times bigger than $C_{use}$ for $g=\{1,2,4,8\}$ respectively, which clearly shows growing preference towards \textbf{use} when the number of infection sources increases.
\end{example}

\subsection{Vaccination: Side-Effect Decision}
\label{sec:side}

Suppose Vaccine $\alpha$ is based on a traditional vaccination technology, hence, it provides protection level $e_\alpha$ with a negligible risk of any undesired side-effect. On the other hand, Vaccine $\beta$ is a product of the most advanced technological improvements, consequently, it offers a stronger protection level $e_\beta$ but with a small likelihood $\epsilon$ of serious undesired consequences. In the following, instead of defining the utility purely based on the vaccine parameters as previously, we utilize explicit costs and benefits variables. $B_\alpha$ and $B_\beta$ are the benefits of the corresponding vaccines. These might differ due to regional diversity of acceptance: one could be accepted worldwide, while the other may be accepted by only specific national authorities. Concerning the costs we capture the cost of infection with $C_i$ while $C_s$ corresponds to the cost of the side-effect which occurs with probability $\epsilon$. We assume the individual is exposed to the virus, hence non-efficient protection correspond to infection.

\begin{definition}
         The Side-Effect decision problem is a tuple $\langle\Sigma,\mathcal{U}\rangle$, where the actions are $\Sigma=\{\alpha,\beta\}$, and the corresponding utility functions $\mathcal{U}=\{U(\alpha),U(\beta)\}$ are presented in Equation (\ref{eq:vacc4}):
\end{definition}

\begin{equation}
    \label{eq:vacc4}
    U(\alpha)=B_\alpha - (1-e_\alpha)\cdot C_i \hspace{2cm}
    U(\beta)=B_\beta - (1-e_\beta)\cdot C_i- \epsilon\cdot C_s
\end{equation}

The optimal decision depends on the exact values of $e_\alpha$, $e_\beta$, $B_\alpha$, $B_\beta$, $C_i$, and $C_s$: if $B_\beta-B_\alpha>\epsilon\cdot C_s-(e_\beta-e_\alpha)\cdot C_i$ then Vaccine $\alpha$ is the optimal choice, otherwise it is Vaccine $\beta$. 
For instance, with $e_\alpha=0.76$, $e_\beta=0.95$, $b_\alpha=b_\beta=100$, $C_i=C_s=1000$, and $\epsilon=0.001$, the utilities are $U(\alpha)\approx-140$ and $U(\beta)\approx49$, respectively. 

\subsection{Proofs}
\label{sec:proof}

\textbf{Proof of Theorem~\ref{th:1}}

    From Table \ref{tab:mask_game_payoffBOTH} it is trivial that both players' cost is minimal when they do not use any masks, i.e., the Nash Equilibrium of the game when both players are susceptible is (\textbf{no}, \textbf{no}). This is also the social optimum, meaning that the players' aggregated cost is minimal. The same holds in case both players are infected, as this only adds a constant $C_i$ to the payoff matrix. 
    
    When only one of the players is susceptible as represented in Table \ref{tab:mask_game_payoffONE}, using no mask is a dominant strategy for the infected player\footnote{Note that the payoffs does not take into account the legal consequences of a deliberate infection such as in \url{https://www.theverge.com/2020/4/7/21211992/coughing-coronavirus-arrest-hiv-public-health-safety-crime-spread}. }, since it is a best response, independently of the susceptible player's action. Consequently, the best option for the susceptible player is \textbf{in}, i.e., the NE is (\textbf{in}, \textbf{no}). On the other hand, the social optimum is different: (\textbf{no}, \textbf{out}) would incur the least burden on the society since $C_{out}<<C_{in}$.

\textbf{Proof of Theorem~\ref{th:2}}

    Independently from the the players actions, the payoff always contains the infection probability multiplied with the cost of infection, hence, we subtract this from the payoffs as shown in Table \ref{tab:util_bayes}. When a players play \textbf{use} her payoff is extended with the cost associated with mask wearing. When neither player plays \textbf{use} the extension consist of the probability of being susceptible while the other is infected multiplied with the infection costs.
    
    (\textbf{use}, \textbf{use}) is dominated by (\textbf{no}, \textbf{use}) and (\textbf{use}, \textbf{no}) as both player would rather play \textbf{no} if the other plays \textbf{use}. Moreover, if the other plays \textbf{no} than they would prefer to play \textbf{no} also if $\rho\cdot(1-\rho)\cdot C_i<C_{use}$. On the other hand, one can easily check that for all values within $[0,1]$ the the expression $\rho\cdot(1-\rho)$ is at most $0.25$ and $C_i>4\cdot C_{use}$ as a consequence of $C_i\gg C_{use}$, hence, the condition does not hold, which makes (\textbf{use}, \textbf{no}) and (\textbf{no}, \textbf{use}) the NEs. These are also social optimums as the total cost is less than the other two cases. 

\textbf{Proof of Theorem~\ref{th:3}}

    From Table \ref{tab:util_eff} it is visible that when a player plays \textbf{no} the other prefers to play \textbf{use} if $C_{use}<a\cdot\hat{C}_i$, which means $\frac{C_{use}}{C_i}<a\cdot\rho\cdot(1-\rho)$. Following the same logic the condition for the player to play \textbf{use} when the other also plays \textbf{use} is $\frac{C_{use}}{C_i}<a\cdot(1-b)\cdot\rho\cdot(1-\rho)$ which is a stronger condition. Consequently, if both is satisfied the players prefer to play \textbf{use}, if neither they play \textbf{no}, and if only the weaker than they play differently.

\textbf{Proof of Theorem~\ref{th:5}}

The strategy vector (\textbf{stay}, \textbf{stay}) is clearly a NE since no player have incentive to deviate from it (because $-C>-\rho\cdot m\cdot L-C$). (\textbf{stay}, \textbf{stay}) could be a NE as well if the same is true, however, the corresponding condition does not hold trivially except when $-C<B-\rho\cdot m\cdot L$ which is equivalent with $\rho\cdot m\cdot L<B+C$. This inequality (if true) also implies that the total payoff is greatest at (\textbf{go}, \textbf{go}), but if it is false than (\textbf{stay}, \textbf{stay}) is the SO. 

\textbf{Proof of Lemma~\ref{th:7}}

{\small
\begin{multline*}
	U_n(p_1,\dots,p_N)=\int_{p_n\cdot t_0}^{\infty} (p_n\cdot B_\beta + (1-p_n)\cdot B_\alpha)\cdot \delta^t \di t -\sum_{i=0}^{n^\prime-1}\int_{\hat{t}_i}^{\hat{t}_{i+1}}C_i\cdot\left(1-\frac{i}{N}\right)\cdot\delta^t \di t = \\
	\left[\frac{(p_n\cdot B_\beta + (1-p_n)\cdot B_\alpha)\cdot \delta^t}{\log(\delta)}\right]_{p_n\cdot t_0}^{\infty} + \sum_{i=0}^{n^\prime-1}\frac{C_i\cdot(N-i)\cdot(\delta^{\hat{t}_i}-\delta^{\hat{t}_{i+1}})}{N\cdot\log(\delta)}= \\
	\left(0-\frac{(p_n\cdot B_\beta + (1-p_n)\cdot B_\alpha)\cdot \delta^{p_n\cdot t_0}}{\log(\delta)}\right) - \frac{C_i}{N\cdot\log(\delta)}\cdot\\
	\left((N-0)\cdot(\delta^{\hat{t}_0}-\delta^{\hat{t}_{1}})+(N-1)\cdot(\delta^{\hat{t}_1}-\delta^{\hat{t}_{2}})+\dots+(N-(n^\prime-1))\cdot(\delta^{\hat{t}_{n^\prime-1}}-\delta^{\hat{t}_{n^\prime}})\right)=\\
	\frac{\delta^{p_n\cdot t_0}\cdot(p_n\cdot B_\beta + (1-p_n)\cdot B_\alpha)}{-\log(\delta)}-\frac{C_i\cdot(N-\delta^{\hat{t}_1}-\dots-\delta^{\hat{t}_{n^\prime}})}{N\cdot\log(\delta)}
\end{multline*}}

\textbf{Proof of Theorem~\ref{th:8}}

If $p_n=p$ than all player's vaccination time is $p\cdot t_0$, so all players gets vaccinated at the same time. Hence, the ordered vaccination times $\hat{t}$ are identical, which implies that no-one get vaccinated before player $n$. As a result, $(N-\delta^{\hat{t}_1}-\dots-\delta^{\hat{t}_{n^\prime}})$ simplifies to $N-\delta^{p\cdot t_0}$.

Since all the players actions are identically $p$, differentiating the utility function in Equation (\ref{eq:av_game2}) with $p$ results in the possible extreme point.\footnote{We should analyse the second order differentiate as well to ensure this extreme point is indeed a maximum, however, since this is merely an example model we do not pursue this analytical direction. } One can check with mathematical software such as \textit{Mathematica} or \textit{Matlab} that the result is indeed as presented in the first point of the theorem. Moreover, by definition $p\in[0,1]$, hence, if this point is outside of this interval the function is monotone, meaning either $p=0$ (i.e., maximally preferring Vaccine $\beta$) or $p=1$ (i.e., maximally preferring Vaccine $\alpha$) is the best response. Indeed, the condition provided in the second and third point of the theorem is the result of comparing the corresponding utility values.
\end{document}